\documentclass[%
 reprint,
 amsmath,amssymb,
 aps,
prl,
longbibliography,
floatfix,
]{revtex4-1}

\usepackage{graphicx}
\usepackage{subfigure}
\usepackage{dcolumn}
\usepackage{bm}
\usepackage[percent]{overpic}

\begin{document}

\preprint{APS/123-QED}

\title{Resonant Very Low- and Ultra Low Frequency Digital Signal Reception Using a Portable Atomic Magnetometer}

\author{Stuart J. Ingleby}
 \email{stuart.ingleby@strath.ac.uk}
\author{Iain C. Chalmers}
\author{Terry E. Dyer}
\author{Paul F. Griffin}
\author{Erling Riis}
\affiliation{%
 Department of Physics, SUPA, Strathclyde University, 107 Rottenrow East, Glasgow, UK\\
}%
\date{\today}

\begin{abstract}
Radio communication through attenuating media necessitates the use of very-low frequency (VLF) and ultra-low frequency (ULF) carrier bands, which are frequently used in underwater and underground communication applications. Quantum sensing techniques can be used to circumvent hard constraints on the size, weight and noise floor of classical signal transducers. In this low-frequency range, an optically pumped atomic sample can be used to detect carrier wave modulation resonant with ground-state Zeeman splitting of alkali atoms. Using a compact, self-calibrating system we demonstrate a resonant atomic transducer for digital data encoded using binary phase- and frequency-keying of resonant carrier waves in the 200~Hz -200~kHz range. We present field trial data showing sensor noise floor, decoded data and received bit error rate, and calculate the projected range of sub-sea communication using this device.
\end{abstract}

\maketitle

\section{Introduction}
Conductive attenuating media, such as seawater, present a longstanding challenge for wireless communication \cite{Moore1967}. Radio communication in media of this type can be achieved using reduced carrier frequencies in the very-low ($<$30~kHz) and ultra-low ($<$3~kHz) frequency bands, exploiting reduced attenuation at lower frequencies. A wide range of activities rely on wireless communication through seawater, and the increasing use of autonomous underwater vessels makes underwater wireless networking desirable \cite{Hattab2013}. However, the high conductivity of seawater remains a serious challenge for radio communication \cite{Meissner2004}, and alternative technologies, such as acoustic or optical communication, are used. However, acoustic communication can suffer from low data rates and increased sensitivity to environmental noise, and optical communication can be of low range and reduced performance in turbid water.

Quantum sensing for communication has been shown to overcome classical constraints on the size and data capacity of signal transducers, based around the principle of resonantly coupling a digitally keyed signal to a quantum system. Recent examples include coupling superconducting qubits to a MHz-frequency resonator \cite{Gely2019}, MHz-carrier phase-shift-keying (PSK) of a magneto-opto-mechanical cavity \cite{Rudd2019}, and reception of amplitude-modulated microwaves using thermal Rydberg atoms \cite{Meyer2018}, demonstrating quantum-enhanced data capacity beyond the classical Chu limit \cite{Cox2018}. 

Radio-frequency atomic magnetometry has been demonstrated with very high sensitivities in unshielded measurements \cite{Keder2014, Deans2018_RSI}, with applications ranging from material defect imaging \cite{Bevington2019} to nuclear quadrupole resonance detection \cite{Cooper2016}. By control of the local static field around an optically pumped sample, the Larmor frequency can be tuned to resonantly detect oscillating fields within a narrow bandwidth, excluding broadband environmental noise. Detection of the resulting atomic spin precession can be performed by measuring the optical rotation of transmitted laser radiation. High signal-to-noise can be achieved using probe laser detuning to maximise the optical rotation signal through non-linear magneto-optical rotation \cite{Gawlik2017}, and homodyne detection of the oscillating signal. We exploit these techniques, and demonstrate a method for calibration of the oscillating signal amplitude to the detected power. Radio-frequency magnetometry can be further enhanced using selective detection of radio-frequency field polarisation \cite{Gerginov2019} and avoidance of back-action through quantum non-demolition measurements \cite{Ciurana2017}.

Previous studies have demonstrated a shielded zero-field atomic magnetometer receiving binary phase-shift-keyed (BPSK) data with carrier frequencies in the range 30-210~Hz \cite{Gerginov2017}. Atomic magnetometry offers two significant advantages over conventional antennae for low-frequency reception: small atomic sensors can replace the large antenna size required for efficient ULF pickup; and the limiting noise sources, such as spin-projection and photonic shot noise, are white, unlike the $1/f$ noise floor imposed by a Johnson-noise limited inductive pickup. In this work we demonstrate a radio-frequency atomic magnetometer, using three-axis compensation to modify the Earth's field and achieve tunable resonant detection and decoding of the keyed carrier. Moreover, these measurements are carried out using portable mass-producible equipment outwith the laboratory.

\section{Double-resonance RF magnetometer}
\begin{figure}
    \centering
    \includegraphics[width=1.0\linewidth]{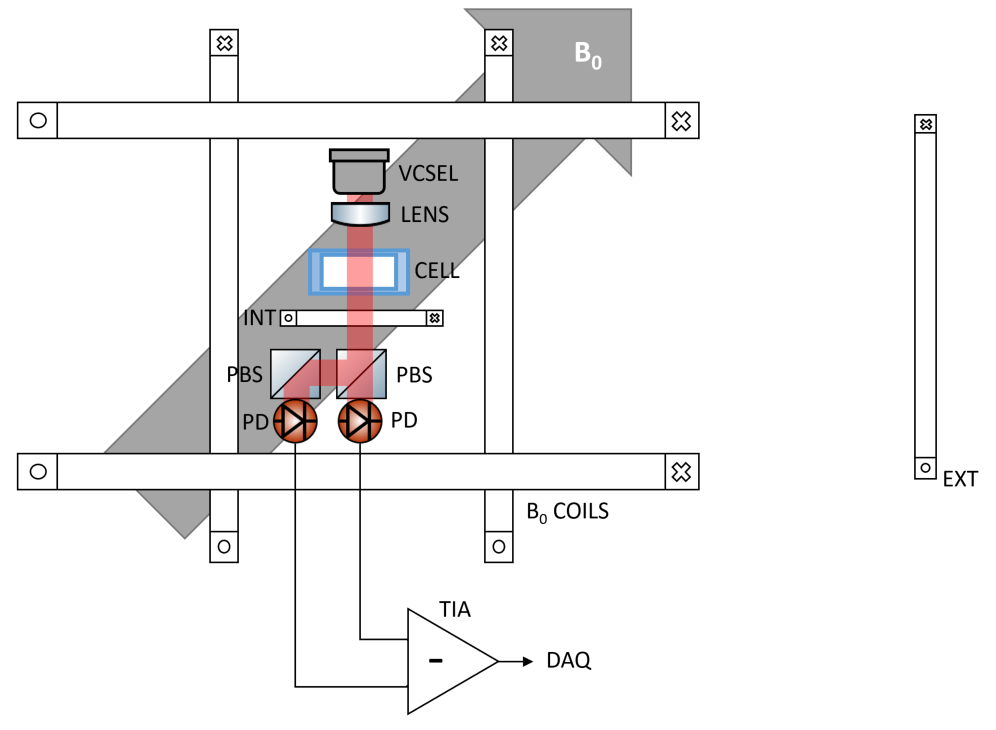}
    \caption{Schematic showing the apparatus used. The double-resonance magnetometer comprises a polarisation-stable vertical cavity surface-emitting laser with integrated electro-thermal feedback (VCSEL), a collimating lens (LENS), a micro-fabricated cell containing a thermal vapour of $^{133}$Cs and approximately 480~torr N$_{2}$ buffer gas, an internal RF modulation coil (INT) and a polarimeter comprising two polarising beam splitters (PBS), photodiodes (PD) and a differential trans-impedance amplifier (TIA). The double-resonance magnetometer is housed in a compact package of 75~cm$^3$ volume. External to this are three-axis static magnetic field coils (B$_{0}$ COILS) surrounding the sensor and an external coil (EXT, not to scale).}
    \label{fig_schematic}
\end{figure}
A compact portable M\textsubscript{x} magnetometer is used, as shown in Figure \ref{fig_schematic}. A vertical-cavity surface-emitting laser (VCSEL) diode, emitting linearly polarised light is used to optically pump and probe a saturated thermal vapour of $^{133}$Cs atoms contained in a microfabricated cell. The VCSEL is thermally stabilised using a thermistor and piezoelectric heater to emit light resonant with the $^{133}$Cs D1 line (894.6~nm), which is collimated to form a beam of diameter 1.4~mm. The VCSEL diode current is provided by a stable DC supply and the VCSEL emits continuously with an output power of 390~$\mu$W. The microfabricated cell consists of a 2~mm thick silicon wafer, photo-lithographically etched to create a cavity of dimensions 5~x~5~x~2~mm and hermetically sealed by anodic bonding of two 0.5~mm thick glass plates on the front and rear surfaces \cite{Hunter2018}. A thermal vapour of $^{133}$Cs and approximately 480~torr N$_{2}$ buffer gas are contained within the cell. The cell undergoes Ohmic heating to approximately 85$^{\circ}$C using a 300~kHz alternating heater current.

Optical pumping of a broad absorption feature, coupled with spin-exchange interactions in the caesium vapour, results in production of a range of ground-state polarisation moments. The VCSEL wavelength is thermally stabilised to maximise the observed amplitude of the first-order vector spin orientation. The precession of this spin orientation is detected by measurement of oscillating birefringence between circular components of the linearly polarised beam, generating polarisation rotation observed at the polarimeter. The polarisation axis of the VCSEL is fixed at 45~degrees relative to the analysis plane of the beam splitters, ensuring that, notwithstanding losses due to optical attenuation or alignment, the polarimeter is approximately balanced.

The evolution of spin orientation under the action of a static field $B_0$ and an oscillating field $B_{RF}$ can be well approximated by considering a total vector magnetisation moment $M$. Solving the equation of motion
\begin{equation}
    \dot{M} = \gamma M \times (B+B_{RF}) - \Gamma M,
\end{equation}
where $\gamma$ is the gyromagnetic ratio for the $^{133}$Cs ground-state and $\Gamma$ is the ground-state spin relaxation rate, assumed here to be constant and isotropic, yields an expression (\ref{eq_resonance})  for the observed signal phase vector detected at the polarimeter.
\begin{equation}
    \begin{pmatrix} \textrm{IP}\\ \textrm{Q} \end{pmatrix} = \frac{A \Omega}{\Gamma^2+\Delta f^2+\Omega^2}\begin{pmatrix} \cos\phi& \sin\phi\\ -\sin\phi & \cos\phi \end{pmatrix} \begin{pmatrix} \Delta f \\ \Gamma \end{pmatrix}
    \label{eq_resonance}
\end{equation}
Equation \ref{eq_resonance} can fitted to demodulated magnetic resonance data, allowing the on-resonance amplitude ($A$) and phase ($\phi$), the magnetic Rabi frequency ($\Omega=\gamma B_{RF}$) and detuning ($\Delta f=f_L - f_{RF}$), and the ground-state spin relaxation rate $\Gamma$ to be estimated. 

\begin{figure}
\includegraphics[width=1.0\linewidth]{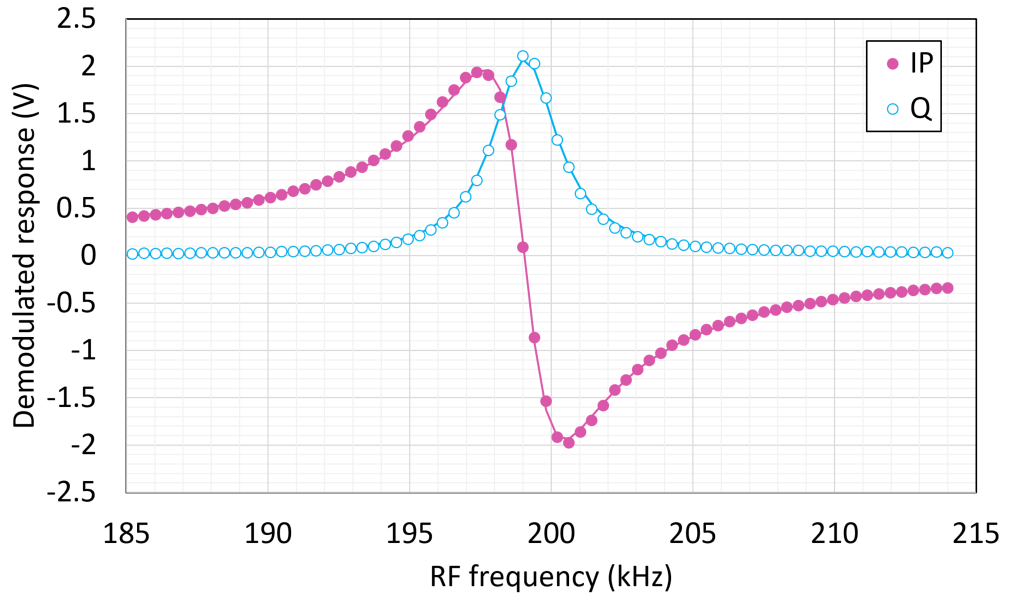}
\caption{Typical magnetic resonance observed with unshielded portable sensor. 75 samples of varying $f_{RF}$ are measured, with each frequency sampled for 20~ms, to average over 50~Hz environmental noise. The in-phase (IP) and quadrature (Q) data points are fitted simultaneously by Equation~\ref{eq_resonance}. Best-fit curves are shown and the resulting fit parameters given in Table~\ref{tab_resonance}. The measured data and fit results are rotated by the fitted on-resonance phase for display.} 
\label{fig_resonance}
\end{figure}

\begingroup
\begin{table}
\caption{Parameters in best fit of Equation \ref{eq_resonance} to the typical magnetic resonance shown in Figure \ref{fig_resonance}.}
\begin{tabular}{ l r l } 
\hline
Parameter & Fit result &\\
\hline
On-resonance amplitude $A$ & 4.627(19) &V \\ 
Larmor frequency $f_0$ & 199.052(9) &kHz \\ 
Spin relaxation rate $\Gamma$ & 749(7) &Hz \\
On-resonance phase $\phi$ & 0.8463(19) &$\pi \cdot$rad \\
Magnetic Rabi frequency $\Omega$ & 1.191(8) &kHz \\
\hline
\end{tabular}
\label{tab_resonance}
\end{table}
\endgroup

Figure \ref{fig_resonance} shows the measured in-phase (IP) and quadrature (Q) components of the polarimeter signal in response to $B_{RF}$ applied to the internal modulation coil over a range of frequencies close to the local Larmor frequency. Fitting Equation \ref{eq_resonance} to these data allows estimation of the Larmor frequency, and hence the static field magnitude $B_0$, and the magnetic Rabi frequency, and hence the oscillating field magnitude $B_{RF}$. This RF frequency scan and fit procedure is software-controlled and can be performed automatically within 2~s. 

In order to tune the Larmor frequency to provide sensitivity to a desired range of RF field frequencies, a three-axis set of software-controlled DC-current coils are used. An iterative calibration process allows full compensation of the geophysical field and $B_0$ to be set in the range 0.25~-~85 $\mu$T at a user-defined orientation. For full details of $B_0$ calibration and control, see \cite{Ingleby2017,Ingleby2018}. The magnetometer sensitivity was optimised for static field orientation at an angle of 45$^{\circ}$ to the laser propagation axis, and this orientation was used throughout the following measurements.

\section{AC field response calibration}

\begin{figure}
\includegraphics[width=1.0\linewidth]{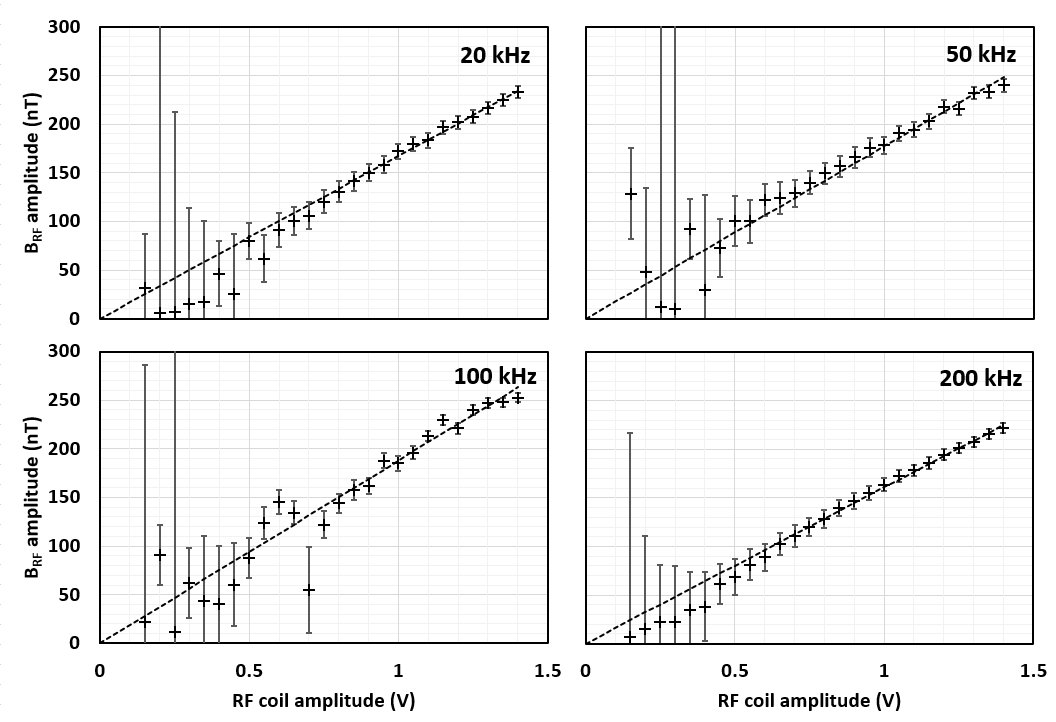}
\caption{Measured internal RF coil calibration. The static field magnitude was varied to set Larmor frequencies in the range 20~-~200~kHz and varying amplitude modulation signals applied to the internal RF coil. Each data point corresponds to $B_{RF}$ inferred from a fit to a magnetic resonance scan. Error bars in $B_{RF}$ correspond to propagated uncertainty in the fit parameters $\Omega$ and $\Gamma$. The resulting response gradients and associated uncertainties are given in Table~\ref{tab_calibration}.} 
\label{fig_calibration}
\end{figure}

Only the component of the RF field which is orthogonal to the static field direction acts to drive the atomic spin precession. For this reason the sensor is calibrated with respect to the observed magnetic Rabi frequency after each adjustment of the static field. Magnetic scans of the type shown in Figure \ref{fig_resonance} were performed for a range of applied amplitudes on the internal modulation coil, and the resulting resonances fitted by Equation \ref{eq_resonance}. The observed magnetic Rabi frequencies allow the amplitude of the detected RF field to be calibrated against the applied internal coil voltage (Figure \ref{fig_calibration}).

\begingroup
\begin{table}
\caption{Measured internal RF coil calibration from weighted linear regression to the data shown in Figure \ref{fig_calibration}. A calculated value is also presented, based on the geometry of the internal modulation coil and alkali vapour cell, assuming a static magnetic field aligned at exactly 45$^{\circ}$ to the light propagation axis.}
\begin{tabular}{ l l } 
\hline
Larmor frequency (kHz) & $B_{RF}$ response (nT/V) \\
 \hline
 20 & 167.2(1.7) \\ 
 50 & 177.3(1.8) \\ 
 100 & 188.0(1.4) \\
 200 & 160.6(1.4) \\
 Calculated & 160.0 \\
 \hline
\end{tabular}
\label{tab_calibration}
\end{table}
\endgroup

The measured magnetic amplitude resulting from applied voltage on the internal coil shows good linearity over the range of RF amplitudes used, is consistent over the range of Larmor frequencies applied, and is in general agreement with the calculated value for an idealised geometry of the coil, cell and static field. This indicates that the inductive load of the coil, which would reduce the response at high frequencies, is negligible for the frequencies used in this work, and that the control of static field orientation is sufficiently precise that the ideal 45$^{\circ}$ angle to the light propagation axis can be achieved repeatably at varying static field magnitudes.

This calibration process requires that the other resonance parameters, such as the ground-state spin relaxation rate $\Gamma$, do not varying for the range of $B_{RF}$ used. Although spin-locking has been observed under increased $B_{RF}$ saturation \cite{Bao2018}, the $B_{RF}$ magnitudes used here are small ($\gamma B_{RF} << \gamma B_0$, and spin-locking effects are not expected to give a $\Gamma(B_{RF})$ dependence.

Under the assumption that the atomic ground-state spin relaxation rate $\Gamma$ is constant in the RF range considered, measurements of the type shown in Figure \ref{fig_calibration} allow the active amplitude of the RF field received by the atomic sensor to be calibrated for the variable geometry and frequency of each subsequent RF field measurement. The calibration procedure takes around 20~s and typically yields a statistical uncertainty of a few percent (see Table \ref{tab_calibration}), which is the dominant source of uncertainty in the calibration of the internal coil. Systematic uncertainties in this calibration include the timing reference used for data acquisition (Keysight 33600A, timebase stability 2 ppm) and the known uncertainty of the gyromagnetic ratio for the $^{133}$Cs F=4 ground-state ($\gamma$ = 3.49862123(35)~Hz/nT \cite{Steck2019}).

\section{Measured AC field response}
Using the static field control and RF response calibration detailed above, a tuned RF magnetic measurement can be made through observation of the polarimeter response around the Larmor frequency, controlled by adjustment of the background static field using the $B_0$ coils. A calibrated modulation at the Larmor frequency applied using the internal coil is detected and demodulated to measure the on-resonance signal response $R(f_L)$ (V/T), and the off-resonance response is scaled by the zero-saturation resonance shape
\begin{equation}
    R(f) = \frac{R(f_L)}{ 1+ \left(\frac{f_L-f}{\Gamma}\right)^2}
    \label{eq_RF_spectrum}
\end{equation}
to account for the bandwidth limit set by the atomic resonance width, assuming small ($\gamma B_{RF} << \Gamma$) detected fields.

\begin{figure}
\includegraphics[width=1.0\linewidth]{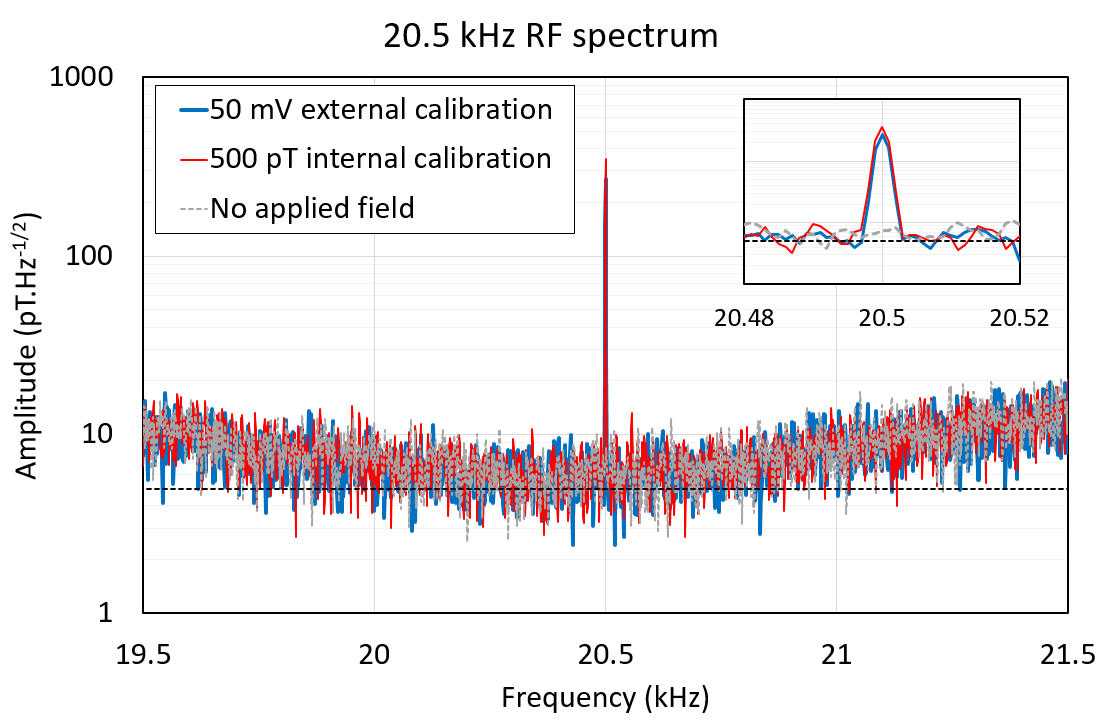}
\caption{Measured RF spectra for a Larmor frequency of 20.5~kHz. Measured polarimeter voltage is calibrated on-resonance using the data in Figure \ref{fig_calibration} and scaled using Equation \ref{eq_RF_spectrum} to account for reduced RF sensitivity off-resonance. Spectra are displayed for a 500~pT amplitude calibration field applied on the internal coil, an uncalibrated 50~mV amplitude field applied on the external coil and no applied field. The uncalibrated external signal is calibrated by demodulation of the internal and external signals. A white noise floor of 5~pT.Hz$^{-1/2}$ is shown for reference. The inset shows a 40~Hz region around the Larmor frequency.} 
\label{fig_20kHz_spectrum}
\end{figure}
Figure \ref{fig_20kHz_spectrum} shows the measured spectra around a Larmor frequency of 20.5~kHz. These data were acquired with a 1~s sample time and 2~MHz sample rate. Spectral response to a 500~pT internal calibration field, a 50~mV external coil field and measured noise floor are shown, and the data is demodulated to determine the amplitude of received signals. Measurements of this type are used to calibrate the received power of the external coil signal. The ratio of observed demodulated amplitudes of the internal and external coil signals, along with the known internal coil calibration allows the external coil amplitude to be estimated. This procedure can be performed rapidly and is used prior to each data transmission measurement. This and the following data was taken using the portable sensor and $B_0$ coils at a quiet rural location, located 1.9~km from the closest main road and 7.7~km from the closest overhead transmission line.

The observed on-resonance noise floor in Figure~\ref{fig_20kHz_spectrum} is close to the  5~pT.Hz$^{-1/2}$ white noise floor shown for reference, and is expected to be limited by environmental pickup and white noise sources such as photon shot noise and atomic spin projection noise. 

To establish the expected limits of this sensor performance, polarimeter signal noise spectra were measured under varying operating conditions. 
\begin{figure}
\includegraphics[width=1.0\linewidth]{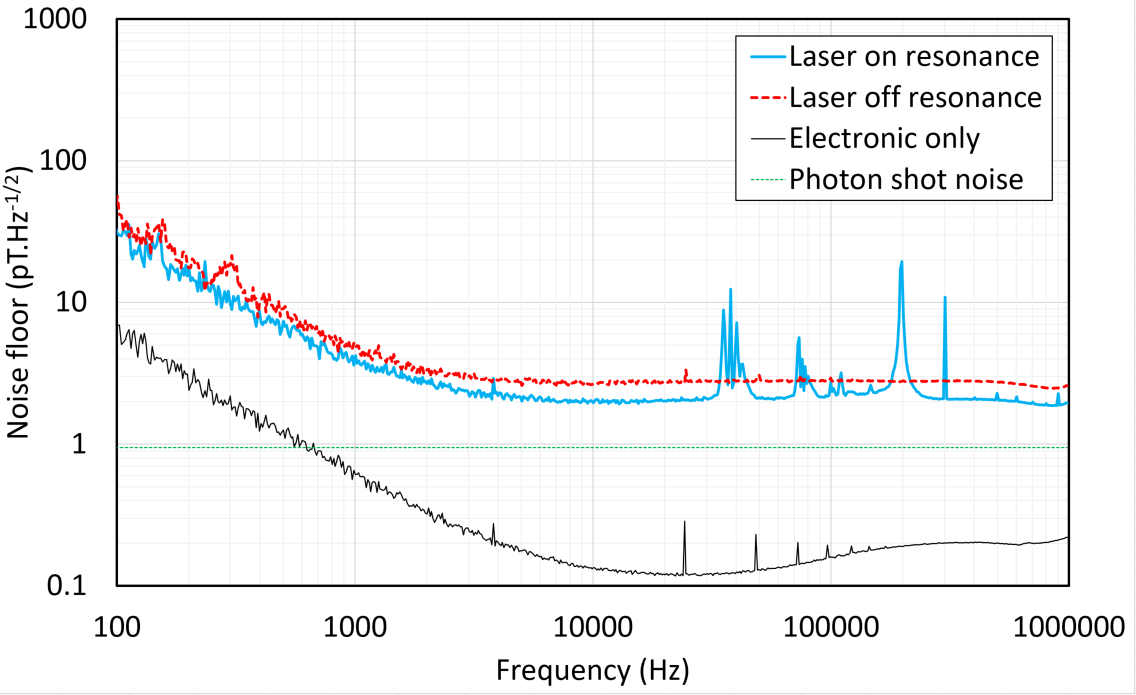}
\caption{Wideband polarimeter spectra, rescaled by the observed RF field calibration. The data shown are the averaged FFT amplitude for 10 samples of 1~s duration, taken for (blue) laser on, tuned to resonance, (red dashed) laser on, red-detuned by approximately 50~GHz and (black) laser off. The green dashed line shows the calculated contribution of photon shot noise to the detected signal.} 
\label{fig_wide_spectrum}
\end{figure}
Figure \ref{fig_wide_spectrum} shows the measured polarimeter spectra, scaled by the calibrated RF field response, for on-resonant light, off-resonant light, electronic noise, and the projected photon shot noise contribution. Variation between the on-resonant and off-resonant polarimeter noise spectra show spurious sources of magnetic pickup, including ambient RF pickup around the Larmor frequency (in this case 200~kHz), the heater current carrier at 300~kHz and spurious heater amplifier noise around 37~kHz and harmonics. However, the increased white noise observed for off-resonant light indicates that the main contribution to the white noise floor is optical, as this is reduced for higher optical absorption on-resonance. This finding is supported by the measurement of electronic noise, which is significantly exceeded by estimated photon shot noise, calculated by
\begin{equation}
    \delta V_{\textrm{PSN}}=G \sqrt{2e S_\lambda P_{\textrm{OPT}}},
\end{equation}
where $G$ is the detector transfer function (V/A), $e$ the electron charge, $S_\lambda$ the spectral sensitivity of the photodiode (A/W) and $P_{\textrm{Opt}}$ the total detected optical power.

\section{Digital data decoding}
The wide tuning range afforded by this RF sensor opens up the possibility of digital data transmission and resonant detection by encoding keyed data in the external RF field. Binary phase-shift keying (BPSK) and frequency-shift keying (FSK) were used to encode alternating data bits onto a carrier frequency $f_c$ close to the sensor's Larmor frequency. In BPSK encoding, symbols are represented by carrier phase shifts of $\pm \pi$~rad, while FSK encodes symbols as binary shifts in frequency $\pm \Delta f_c$. Signal recovery and decoding was demonstrated for RF data transmitted from the external coil at varying carrier frequencies 200~kHz - 200~Hz for digital symbol rates $f_s$ ranging between 100~bit/s - 1~kbit/s. For the FSK data presented here the FSK shift was equal to the observed resonance width $\Delta f_c = \Gamma$.

The digitally-keyed modulated RF signal was transmitted to the portable sensor using a 30~cm diameter 15~turn coil (the external coil in Figure \ref{fig_schematic}) driven by an arbitrary function generator (Keysight 33600A). At each carrier frequency the sensor static field was tuned to set the Larmor frequency to the RF carrier frequency, the sensor response to the internal modulation coil calibrated, and the demodulated amplitudes of the calibrated internal signal and received (external) carrier signal used to measure the received carrier amplitude. The received carrier power was calculated using
\begin{equation}
    P_c=\frac{c}{\mu_0} a B_c^2
\end{equation}
where $a$ is the transverse area of the probed atomic sample, estimated using the dimensions of the microfabricated alkali cell and the collimated laser beam, and $B_c$ is the detected RF carrier signal amplitude.
\begin{figure}
\begin{tabular}{c}
\includegraphics[width=1.0\linewidth]{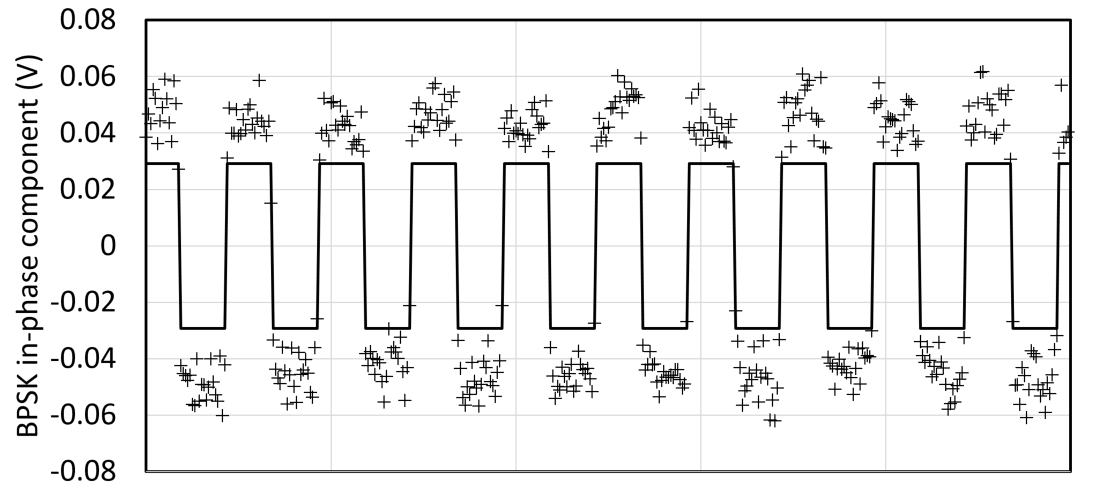} \\
\includegraphics[width=1.0\linewidth]{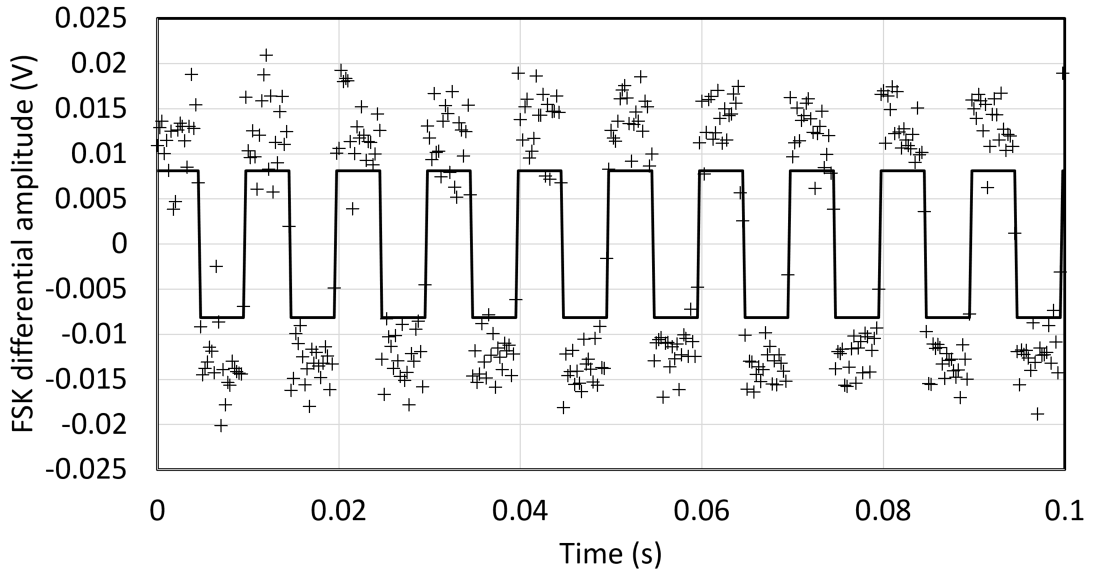}
\end{tabular}
\caption{Received digital data at 200~bit/s modulating a 5~kHz carrier, sampled at 2~MHz, demodulated and down-sampled to 1~kHz. Received carrier power is -60~dBm. No erroneous bits are observed in this sample. } 
\label{fig_5kHz_data}
\end{figure}
The received signal is sampled at 2~MHz and decoded. For BPSK the raw data is demodulated at the carrier frequency, then the signal phase demodulated at the symbol rate $f_s$. For FSK keying the raw data is coherently demodulated at both $f_c+\frac{\Delta f_s}{2}$ and $f_c-\frac{\Delta f_s}{2}$, and the resulting differential amplitude demodulated at $f_s$. In both cases the symbol phase is recovered and the data low-pass filtered and down-sampled to $f_{\textrm{DS}}$. Figure \ref{fig_5kHz_data} shows down-sampled data for BPSK and FSK data received with a signal power of -60~dBm, carrier frequency of 5~kHz and data rate of 200~bit/s. 

The bit error rate (BER), defined as the ratio of erroneous bits $N_e$ to the total number of transmitted bits $N$, is determined at each transmission frequency and power by demodulation of $N$ consecutive bits. In the presence of additive white Gaussian (AWG) noise, bit error probabilities for BPSK and coherent FSK demodulation can be estimated by
\begin{equation}
    \textrm{BER}_{\textrm{BPSK}}=\frac{1}{2}\left(1-\textrm{erf}\left(\sqrt{\frac{E_b}{N_0}}\right)\right)
\end{equation}
\begin{equation}
    \textrm{BER}_{\textrm{FSK}}=\frac{1}{2}\left(1-\textrm{erf}\left(\sqrt{\frac{E_b}{2N_0}}\right)\right)
\end{equation}
where $E_b$ is the signal energy received per bit and $N_0$ is the AWG noise energy.
\begin{figure}
\begin{tabular}{c}
\includegraphics[width=1.0\linewidth]{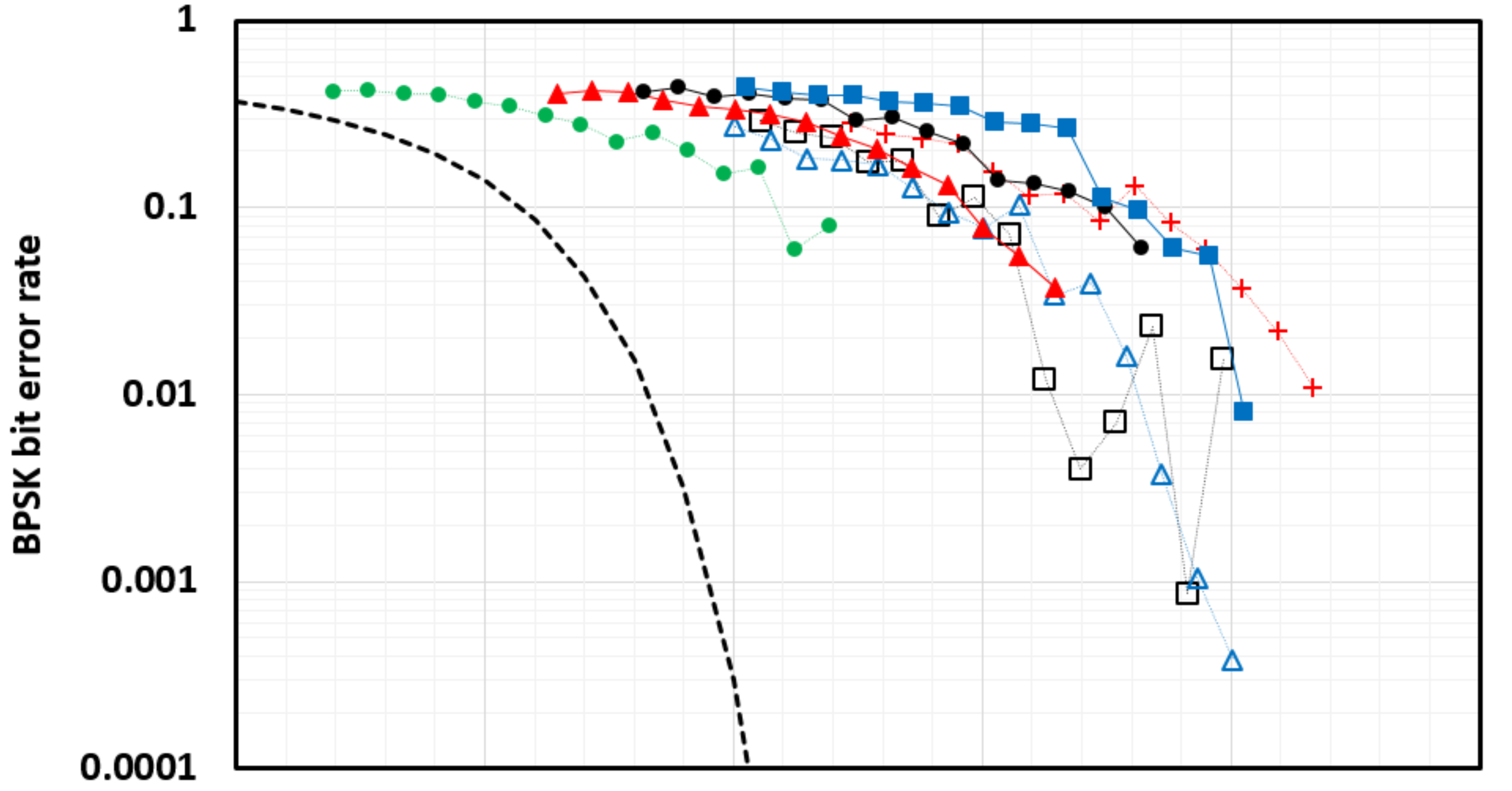} \\
\includegraphics[width=1.0\linewidth]{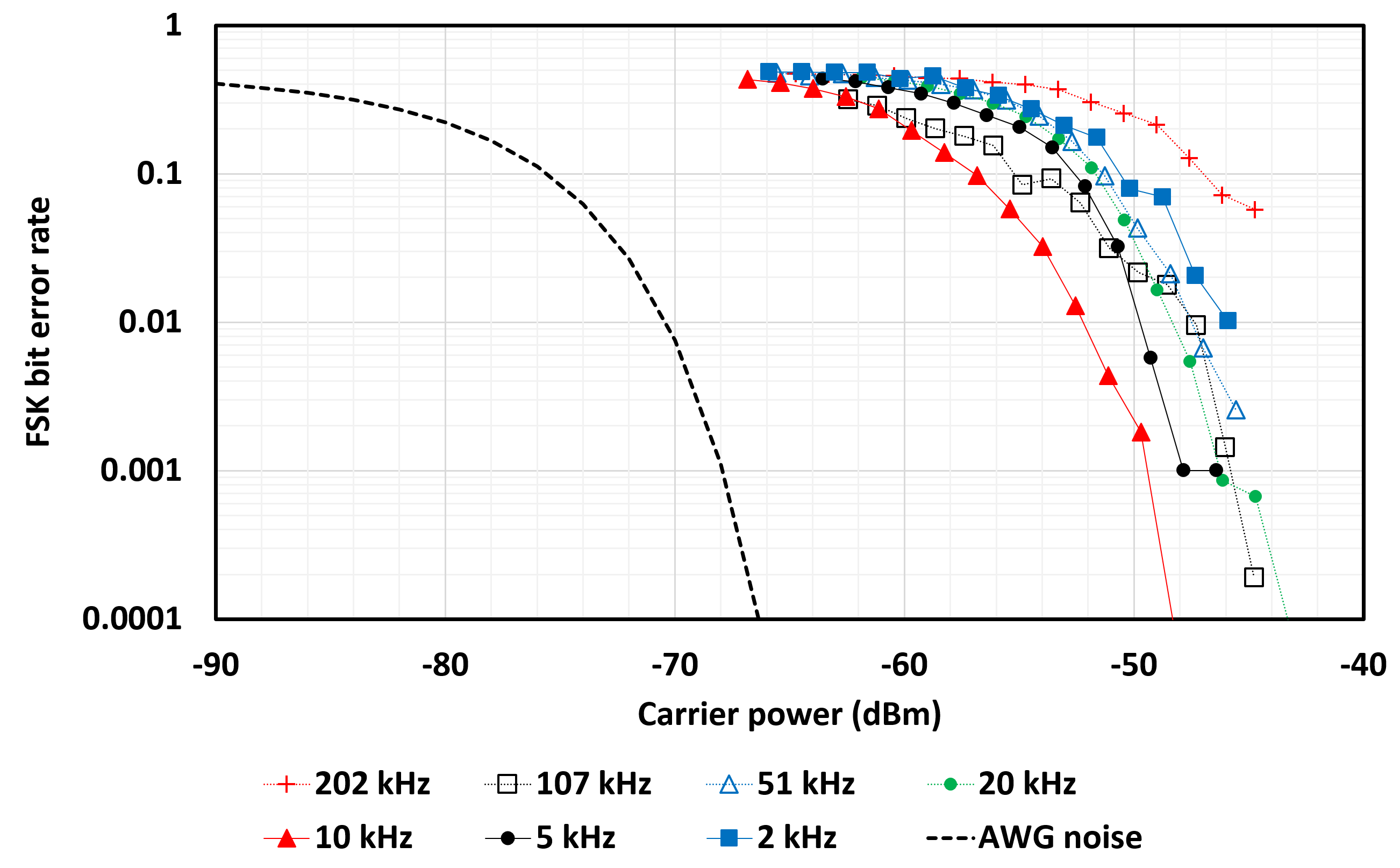}
\end{tabular}
\caption{Measured bit error rates for 1~kbit/s (top) BPSK and (below) FSK keyed data over a wide range of carrier signal power and frequency. Dashed lines between the data points are included to guide the eye, and the minimum bit error rate in the presence of additive white Gaussian noise of 5~pT.Hz$^{-\frac{1}{2}}$ is shown (black dashed line). N~=~$10^4$~bits were received at each carrier frequency and power, resulting in a 1\% statistical uncertainty in the measured bit error rate.} 
\label{fig_1kbps_data}
\end{figure}
Data reception and decoding was tested at $f_s=$ 1~kbit/s for both BPSK and FSK for a range of carrier frequencies 2-200~kHz and the bit error rate measured (Figure \ref{fig_1kbps_data}). 

\begin{figure}
\begin{tabular}{c}
\includegraphics[width=1.0\linewidth]{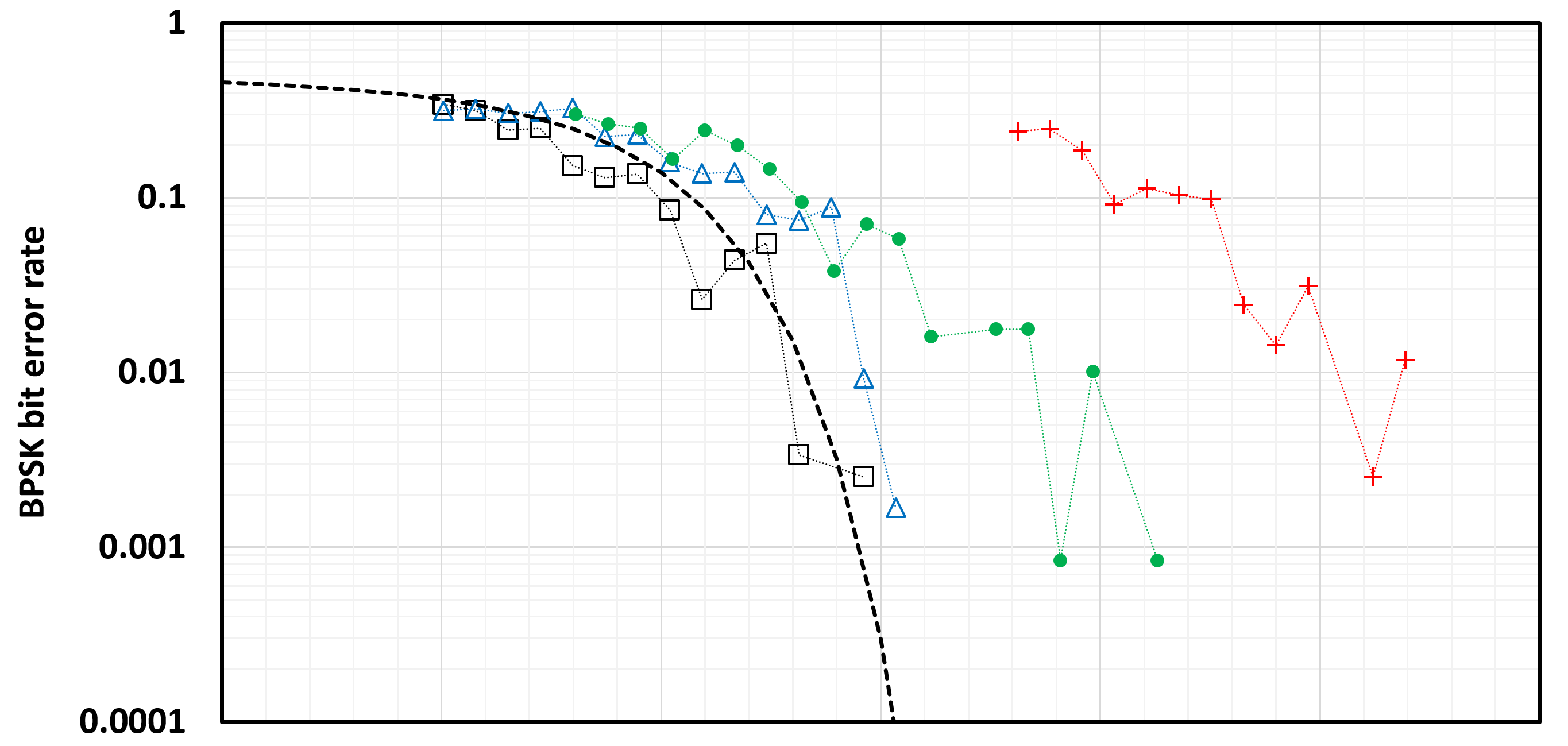} \\
\includegraphics[width=1.0\linewidth]{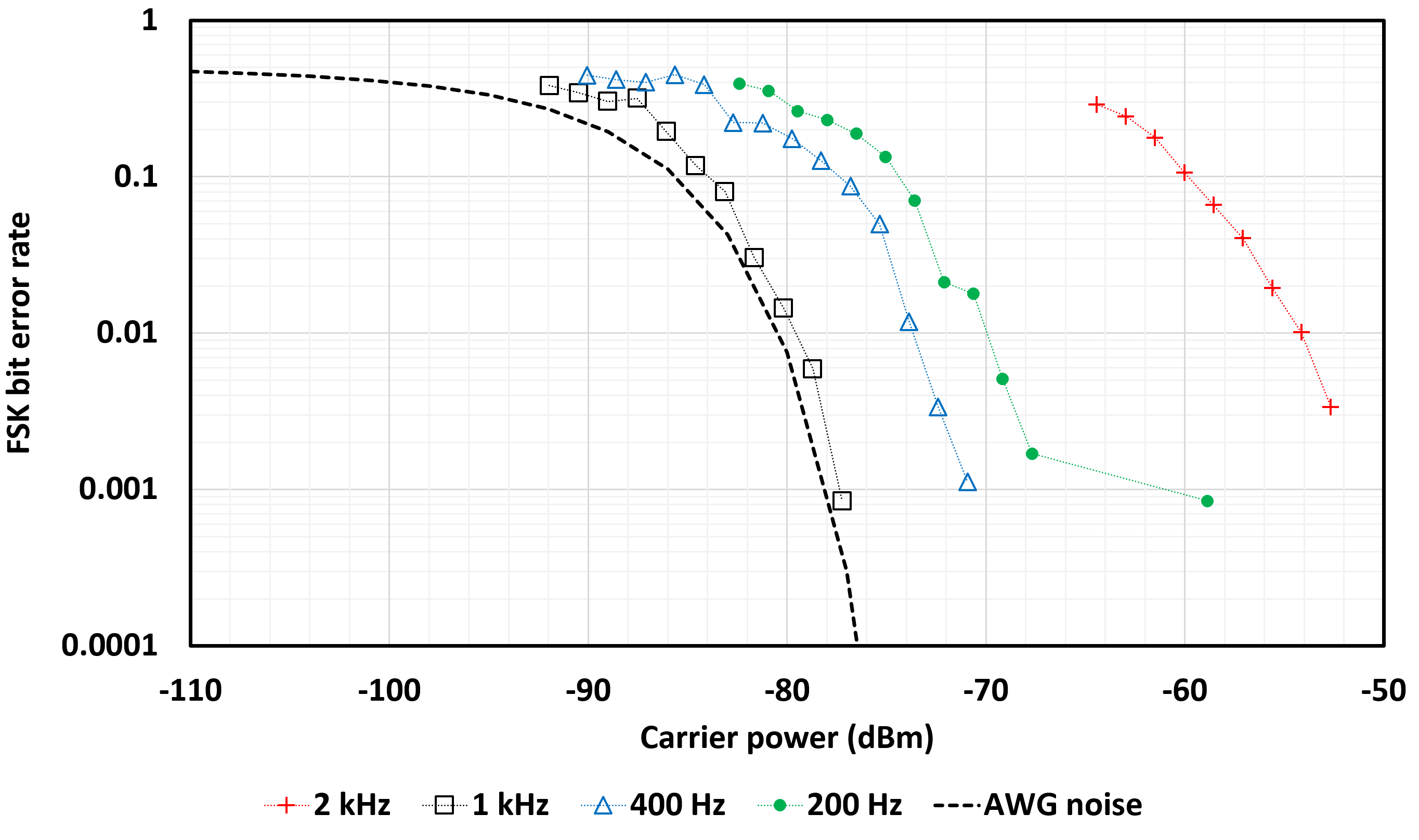}
\end{tabular}
\caption{Measured bit error rates for (top) BPSK and (below) FSK keyed data at 100~bit/s over a wide range of carrier signal power and frequency. Dashed lines between the data points are included to guide the eye, and the minimum bit error rate in the presence of additive white Gaussian noise of 5~pT.Hz$^{-\frac{1}{2}}$ is shown (black dashed line). N~=~1111~bits were received at each carrier frequency and power, resulting in a 3\% statistical uncertainty in the measured bit error rate.} 
\label{fig_100bps_data}
\end{figure}

In order to reduce the carrier frequency further and access the ultra-low and super-low frequency regimes, the data rate was then reduced to 100~bit/s and bit error rate measured for carrier frequencies of 0.2-2~kHz (Figure \ref{fig_100bps_data}).

\begin{figure}
\begin{tabular}{c}
\includegraphics[width=1.0\linewidth]{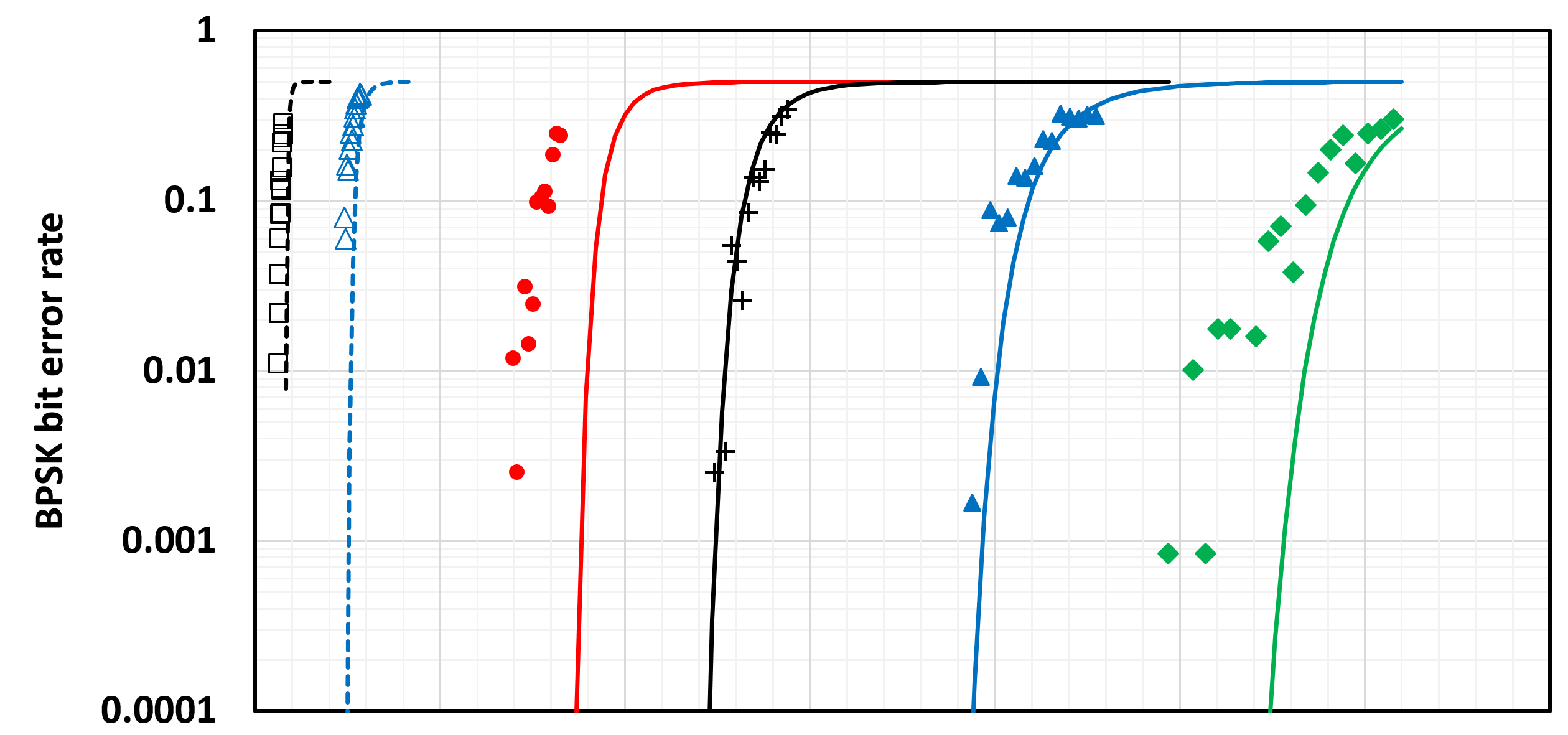} \\
\includegraphics[width=1.0\linewidth]{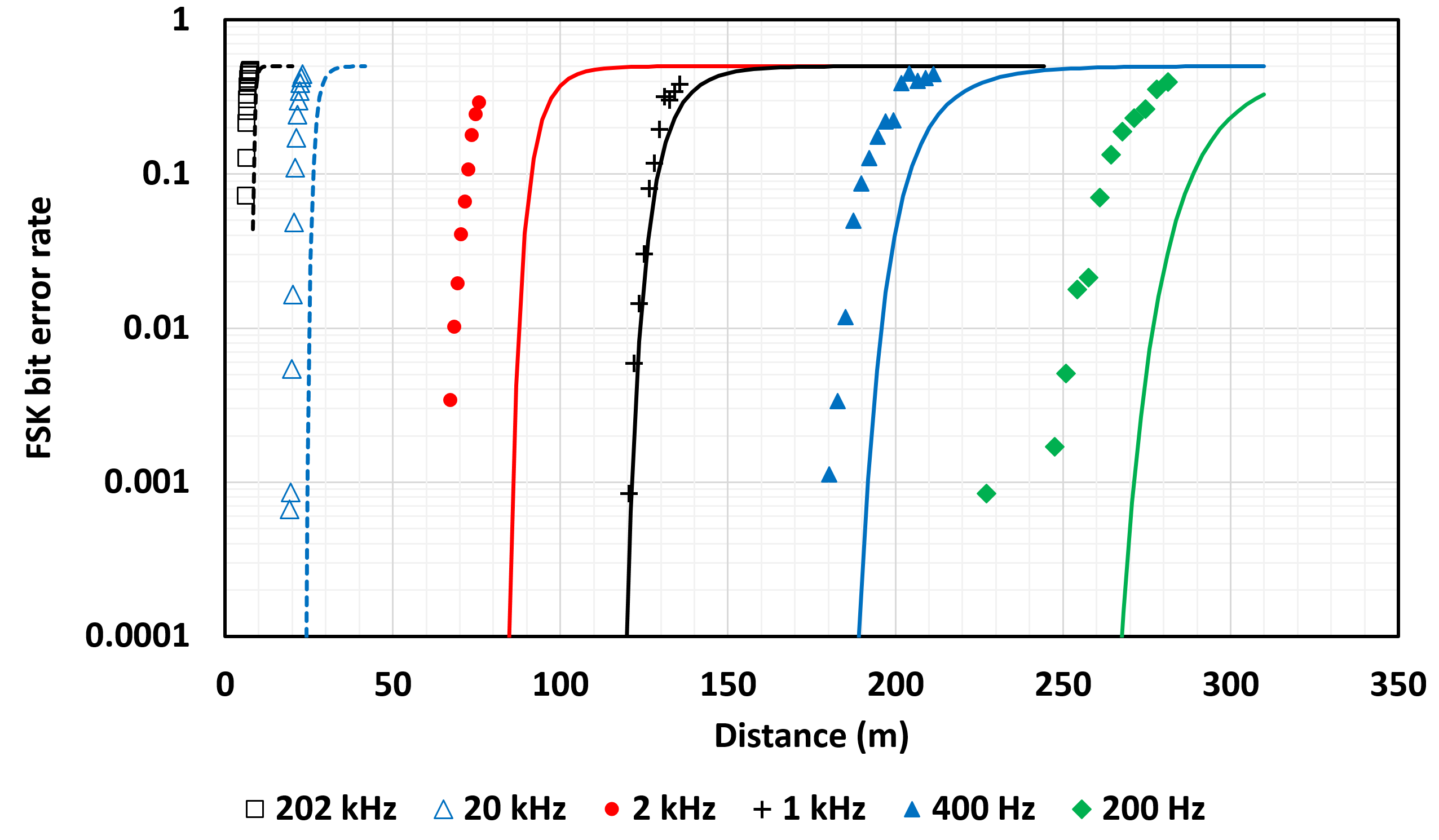}
\end{tabular}
\caption{Measured BER rescaled to estimated seawater attenuation for (top) BPSK and (bottom) FSK keyed data transmitted at 100~bit/s with a carrier power of 10~W. Lines show the minimum bit error rate in the presence of additive white Gaussian noise of 5~pT.Hz$^{-\frac{1}{2}}$ at each carrier frequency. Dashed lines and hollow markers denote 1~kbit/s data. } 
\label{fig_range}
\end{figure}

In order to quantify the utility of lower carrier frequencies for data transmission in attenuating media, the data shown in Figures \ref{fig_1kbps_data} - \ref{fig_100bps_data} can be used to calculate the attenuation length of a 10~W carrier transmitter in seawater. The attenuation constant of seawater has been reported in \cite{Hattab2013}, and can be approximated, for water temperature of 17$^\circ$C, below 100~MHz by $\sqrt{f_c}/32.5$~dB/m, where $f_c$ is the carrier frequency in Hz. Figure \ref{fig_range} shows the measured BER data rescaled to the calculated distance at which a  10~W transmitted carrier would be reduced to the detected carrier power in the absence of dispersion and other path losses. 

\section{Discussion}
The results presented demonstrate the performance of this compact portable sensor. The approximate noise floor of 5~pT.Hz$^{-1/2}$ does not yet exceed conventional inductive sensors \cite{Hospodarsky2016}. The feature observed in Figure \ref{fig_wide_spectrum} at the Larmor frequency demonstrates that there is significant resonant magnetic pickup in the absence of applied internal or external RF fields, which may be environmental noise or represent interference at the sensor due to other parts of the measurement system, such as the data acquisition unit or external function generator. The sensor's electronic noise floor is exceeded by the observed photonic noise floor (approximately 2~pT.Hz$^{-1/2}$). The reported system differs from sensors such as those reported in \cite{Deans2018_RSI,Keder2014,Bevington2019} in our use of a VCSEL and microfabricated vapour cell. These design features, chosen to enhance the system's readiness for portable applications, are subsystems for further optimisation in order to reduce sensor noise towards these best-in-field examples.

Noise floor aside, the sensor's widely tunable resonant atomic response represents an advantage over inductive sensors. The resonance frequency may be modified quickly over a three-decade range without any hardware changes, and the resonant atomic response suppresses off-resonant interference prior to any additional filtering. These features would be suited to applications requiring frequency hopping or reception of low signal-to-noise data.

The sensor also benefits from an SI-traceable calibration of received RF power. Calibration of the reported data is statistically limited, but in applications in which absolute measurement of LF/VLF/ULF fields is required, the use of better timing sources and longer calibration measurement would result in calibration tolerances limited by known uncertainties for alkali ground-state gyromagnetic ratios in the linear Zeeman regime. 

For carrier frequencies ranging from 200~Hz to 200~kHz, bit error rates were observed to fall below 10\% for received carrier powers not exceeding -40~dBm (10$^{-7}$~W), and the expected roll-off of BER with carrier power was observed. However, the projected minimum BER in the presence of AWG noise at the sensor noise floor was not achieved over the full range of carrier and symbol rates. Significant variation in BER for both BPSK and FSK is observed at different carrier frequencies, without strong correlation to the relative change in frequency. This may be indicative of the dominant interference being non-white or non-Gaussian, for example, harmonic background interference. The consistent excess BER observed for data transmitted with a symbol rate of 1~kbit/s (Figure \ref{fig_1kbps_data}) reflects the clipping of the modulation signal spectrum at 1~kbit/s by the sensor bandwidth, given by the atomic relaxation rate $\Gamma$. Nevertheless the authors thought that inclusion of kbit/s data transfer results was useful to the reader.

\section{Conclusions}

We report measured RF signal reception and decoding for a portable resonant atomic sensor in the LF, VLF, ULF and SLF communication bands. The sensor's response to RF signals is calibrated with respect to the measured saturation of $^{133}$Cs ground-state Zeeman transitions and a noise floor of 5~pT.Hz$^{-1/2}$ observed in non-laboratory field trials. Reception of BPSK and FSK keyed data over carrier frequencies in the range 200~Hz to 200~kHz is demonstrated and received bit error rates measured for carrier powers ranging between $-40$ and $-100$~dBm. Comparison with theoretically allowed bit error rates is made and attenuation ranges for sub-sea communications estimated.

\section{Acknowledgements}
This work was funded by the UK Quantum Technology Hub in Sensing and Metrology, EPSRC (EP/M013294/1). The data shown in this paper is available for download at TBC. The authors would like to thank Mr David Upton and the staff at Ross Priory for their assistance in organising the reported field trials.

\bibliography{main.bib}
\end{document}